\documentclass[debug]{rmaa}
\newcommand{\Msun}{\mbox{$\rm M_{\odot}$}}
\newcommand{\Rsun}{\mbox{$\rm R_{\odot}$}}
\newcommand{\Lsun}{\mbox{$\rm L_{\odot}$}}
\usepackage{paralist}

\usepackage[latin1]{inputenc}

\title{The Absolute Parameters for NSVS\,11868841\thanks{Based on spectroscopic 
observations collected at T\"{U}B\.{I}TAK National Observatory (Turkey).} and 
the Oversized Stars in the Low-Mass Eclipsing Binaries
} 

\author{
  \"{O}. \c{C}ak{\i}rl{\i},\altaffilmark{1,2} 
  C.~\.{I}bano\v{g}lu,\altaffilmark{1}
  and A. Dervi{\c s}o{\v g}lu\altaffilmark{1}}

\altaffiltext{1}{Ege University, Science Faculty, Astronomy and Space Sciences Dept., 35100 Bornova, \.{I}zmir, Turkey. $e-mail$: omur.cakirli@ege.edu.tr}
\altaffiltext{2}{T{\"U}B{\.I}TAK National Observatory, Akdeniz University Campus, 07058 Antalya, TURKEY}

\shortauthor{\c{C}akirli, \.{I}bano\v{g}lu, \& Dervi{\c s}o{\v g}lu}
\shorttitle{The Absolute Parameters for NSVS\,11868841}

\fulladdresses{
\item Ege University, Science Faculty, Astronomy and Space Sciences Dept., 35100 
Bornova, \.{I}zmir, Turkey. $e-mail$: omur.cakirli@ege.edu.tr}

\listofauthors{\"{O}. \c{C}ak{\i}rl{\i}, C. \.{I}bano\v{g}lu, \& A. Dervi{\c s}o{\v g}lu}
\indexauthor{\c{C}ak{\i}rl{\i}, \"{O}}
\indexauthor{\.{I}bano\v{g}lu, C.}
\indexauthor{Dervi{\c s}o{\v g}lu, A.}

\abstract{Spectroscopic observations of the low-mass eclipsing binary NSVS\,11868841\ have 
been obtained and the radial velocities were derived for both components.
The masses and radii determined for the components are  
$M_1$ =0.870$\pm$0.074\Msun, $M_2$=0.607$\pm$0.053\Msun and $R_1$ =0.983
$\pm$0.030 \Rsun, $R_2$ =0.901$\pm$0.026 \Rsun.  Both the primary and secondary 
stars' radii are 10 \% and 57\% larger than those of zero-age-main-sequence stars with the 
same masses. This discrepancy may be arisen from the large spot coverage of both stars.
We collected absolute parameters of 21 low 
mass double-lined eclipsing binaries and compared their positions  in the mass-radius and mass-effective 
temperature panels.  The large radii and lower effective temperatures are solved neither with difference in 
metallicity nor in mixing length parameters. These discrepancies  in the low mass stars may be originated 
by magnetic fields causing 
inhibition convective energy transport which leads to large magnetic spot coverage on the surface of a low 
mass star.
}

\addkeyword{Stars: activity}
\addkeyword{Stars: fundamental parameters}
\addkeyword{Stars: low-mass}
\addkeyword{Stars: eclipsing}

\begin{document}
\maketitle

\section{Introduction}
\label{sec:intro}
Eclipsing binaries are the best laboratories for determining the fundamental physical properties of stars. Detached, 
double-lined eclipsing binaries yield direct and accurate measures of the masses and radii of the component stars. 
Measuring of these quantities has always been observationally 
challenging. Large-scale surveys, providing imaging and photometric catalogues for substantial fractions of the 
celestial sphere, are now available at a wide range of wavelengths. Over the last few years, these surveys have 
become invaluable research tools for investigating the properties of intrinsically rare objects. These surveys 
have proven most useful in searches for low-mass stars. Stars with masses less than 1 \Msun~are regarded as 
low-mass stars  and have late spectral types, namely G, K or M. We cannot vary the conditions prevalent on an 
individual star, so observations of many individual stars are combined to build up a picture of how stars of 
different mass and composition evolve. Since the  intermediate- and high-mass stars are intrinsically bright sources 
most studies have centered on them with 
little consideration of M dwarfs with masses below 0.6 \Msun. This reflects the availability of more precise 
observational constraints, and also greater analytic tractability in modelling higher-mass stars. Recently, however, 
the lower main-sequence has attracted more attention, with a series of detailed models extending past the 
hydrogen-burning limit to the boundary between low-mass brown dwarfs and giant planets. Understanding the 
structure and evolution of stars is a basic goal of stellar astronomy. Critical tests of the evolution theory 
for stars can be made on a small set of eclipsing binary stars (Baraffe et al. 1998, Girardi et al. 2002, 
Siess et al. 1997, Chabrier et al.  2007). While the results of stellar evolution models compare favourably 
to data for main-sequence stars with masses greater than that of the Sun, evidence has been growing that 
the models for stars on the lower main-sequence have problems when confronted with precise masses and radii 
from double-lined eclipsing binaries.  As shown by Torres and Ribas (2002) the evolutionary models 
underestimate the radii  and overestimate the effective temperatures of the low-mass stars. These discrepancies 
were confirmed in subsequent studies  by Ribas(2003), Lopez-Morales and Ribas (2005), Torres et al. (2006), 
Morales et al. (2009a), Morales et al. (2009b), and others. Hints of the radius discrepancies have also been described by 
Popper (1997), Clausen et al. (1999),  and others. Recently,  Mullan and MacDonald (2001),  Torres et al. (2006), 
Morales, Ribas and Jordi (2008) proposed a hypothesis based 
upon the effects of stellar activity to explain the discrepancies. Furthermore Stassun et al. (2009) called 
attention to that low-mass stars appear older or younger in mass-radius diagram depending on whether 
post- or pre-main-sequence models are used. As the number of low-mass eclipsing binaries studied carefully, both 
photometrically and spectroscopically, is increased a better comparison with the theoretical models can be made. 
   
Light variability of  NSVS\,11868841 ( $\alpha =$ 23$^h$17$^m$58$^s$,\, $\delta =$ + 19\arcdeg 17\arcmin 03\arcsec
(J2000), hereafter NSVS\,1186) was detected by Wozniak et al.(2004) from the 
{\sc Northern Sky Variability Survey}. They classified it as a detached eclipsing binary with combined median  
magnitude  of V=14$^m$.084$\pm$0.076 and an orbital period of 0.60179 days. High-precision light curves 
in the Johnson V-, R-, and I-bandpass  were obtained by Coughlin and Shaw (2007). Since the radial velocities 
of the components were not available they have determined the radii of the stars by analysing the V- and 
R-bandpass light curves. In this study we present the first radial velocities for the components of 
double-lined, short-period eclipsing binary NSVS\,1186 and the physical parameters obtained from the analyses 
of radial velocities and light curves. Both components show high level chromospheric  activity, evidenced 
by H$_{\alpha}$ emission lines in their spectra. We discuss the properties of the binary and in relation to 
the discrepancies between empirical and theoretical mass-radius and mass-effective temperature relations 
for low-mass stars.

\section{Spectroscopic Observations}
\subsection{Spectroscopy}
Since the wide-band light curves of NSVS\,1186 are available we observed the star spectroscopically. Optical 
spectroscopic observations were obtained with the Turkish Faint Object Spectrograph Camera (TFOSC) attached 
to the 1.5 m telescope on  August 23, 24, and 25, 2008 under good seeing conditions. Further details on the 
telescope and the spectrograph can be found at \url{http://www.tug.tubitak.gov.tr}. The wavelength coverage of each 
spectrum was 4100-8100 \AA~ in 11 orders, with a resolving power of $\lambda$/$\Delta \lambda$ 7\,000 at 
6563 \AA~ and an average signal-to-noise ratio (S/N) was $\sim$150. We also obtained a high S/N spectrum 
of the M dwarf GJ\,740 (M0 V) and GJ\,623 (M1.5 V) for use as templates in derivation of the radial 
velocities (Nidever et al. 2002). 

The electronic bias was removed from each image and we used the 'crreject' option for cosmic ray removal. The echelle spectra were extracted and wavelength calibrated by using a Fe-Ar lamp source with help of the 
IRAF {\sc echelle} package. The stability of the instrument was checked by cross correlating the spectra of the standard star against each other using the {\sc fxcor} task in IRAF. The standard deviation of the differences 
between the velocities measured using {\sc fxcor} and the velocities in Nidever et al. (2002) was 
about 1.1 km s$^{-1}$.

\subsubsection{Spectral Types and Temperature Estimates}
We have used the spectra to reveal the spectral type of the primary component of NSVS\,1186. For this 
purpose we have degraded the spectral resolution from 7\,000 to 3\,000, in order to use the calibrations
 by Hern\'andez et al. (2004) by convolving them with a 
Gaussian kernel of the appropriate width, and we have measured the equivalent widths ($EW$) of 
photospheric absorption lines for the spectral classification. We have followed the procedures 
described by Hern\'andez et al., choosing metallic lines in the blue-wavelength region, where 
the contribution of the primary component is considerably larger than that of the secondary in the 
spectra. From several spectra we measured equivalent widths of some spectral lines assuming the light contribution of the primary star is 0.6, as given in Table\,1.

\begin{table}
\begin{center}
\caption{Equivalent widths of the selected lines in the spectra.}
\label{Table 1.}
\begin{tabular}{ccc}
\hline
 $Spectral~lines$ & EW$_{p}$ (\AA)  & EW$_{s}$ (\AA)  \\
\hline
Ca \sc i $\lambda$ 4226			&2.13$\pm$0.11 	& 1.39$\pm$0.12	\\
Fe \sc i $\lambda$ 5329			&1.11$\pm$0.11 	& 0.99$\pm$0.12	\\
Fe \sc i $\lambda$ 4271			&1.44$\pm$0.13 	& 1.41$\pm$0.12	\\
Ca {\sc i}+Fe {\sc i} $\lambda$ 5270	&2.91$\pm$0.21 	& 2.11$\pm$0.21	\\
Ca \sc i $\lambda$ 6162			&1.99$\pm$0.42 	& 1.11$\pm$0.09	\\
Fe \sc i $\lambda$ 5079  		&0.79$\pm$0.07 	& 0.47$\pm$0.11	\\
\hline \\
\end{tabular}
\end{center}
\end{table}

From the calibration relations of $EW$--Spectral-type given by Hern\'andez et al. (2004), we have derived a 
spectral type of G9$\pm$1 for the primary  component. The effective temperature 
deduced from the calibrations of Drilling \& Landolt (2000), de Jager \& Nieuwenhuijzen (1987) and Popper (1980) 
is 5\,230$\pm$80, 5\,230$\pm$80, and 5\,330$\pm$141\,K for the primary component, respectively. 
 The mean effective temperature of the primary component deduced from the spectra is 5\,250$\pm$95 \,K .

NSVS\,1186 is listed in several large photometric databases consolidated in the {\sc Naval Observatory Merged 
Astronomical Dataset} (NOMAD-1.0, Zacharias et al. 2004), which provide optical magnitudes of 
B=14$^m$.77$\pm$0$^m$.02, V=13$^m$.87$\pm$0$^m$.01, R=13$^m$.94$\pm$0$^m$.01,
 I=12$^m$.74$\pm$0$^m$.02.Since the magnitudes collected from photographic measurements  
 and the colors are inconsistent no attempt has been made to calculate the effective temperatures 
 of the components. The infrared magnitudes are taken from 2MASS (Cutri et al. 2003) catalog as 
J=12$^m$.518$\pm$0$^m$.023, H=12$^m$.069$\pm$0$^m$.021, and K=11$^m$.995$\pm$0$^m$.019.
The observed infrared colours of J-H=0$^m$.449$\pm$0$^m$.031 and H-K=0$^m$.074$\pm$0$^m$.028  
correspond to a combined spectral type of G9$\pm$2 is in a good agreement with that we derived by spectral lines 
alone. Using the depth of the eclipses in three bands we estimate light contribution  of the primary component as 0.60,
0.57, 0.55 and 0.56 in the V, J, H and K bands respectively. Hence V-K, J-H and H-K colors of the primary component corresponds
to a spectral type of G9$\pm$2. We estimated a temperature of 5\,240\, $\pm$230\,K from the calibrations 
of Tokunaga (2000). Temperature uncertainty of the primary component results from considerations
 of spectral type uncertainties, and calibration differences. The effective temperature of the primary star
  what we derived from the infrared photometric measurements is in a good agreement with that 
  we estimated from the spectra alone. The weighted mean of the effective temperature
   of the primary component is 5\,250$\pm$135\,K.

\subsubsection{Rotational Velocities of the Components}
The width of the cross-correlation profile is a good tool for the measurement of $v \sin i$ (see, e.g., 
Queloz et al. 1998). The rotational velocities ($v \sin i$) of the two components were obtained by 
measuring the FWHM of the CCF peaks in nine high-S/N spectra of NSVS\,1186 acquired close to the 
quadratures, where the spectral lines have the largest Doppler-shifts. In order to construct a 
calibration curve FWHM-$v \sin i$, we have used an average spectrum of HD\,27962, acquired with 
the same instrumentation. Since the rotational velocity of HD\,27962 is very low but not zero 
($v \sin i$ $\simeq$11 km s$^{-1}$, e.g., Royer, Zorec \& Fremat 2004 and references therein), it 
could be considered as a useful template for low-mass stars rotating faster than $v \sin i$ $\simeq$ 
10 km s$^{-1}$. The spectrum of HD\,27962 was synthetically broadened by convolution with rotational 
profiles of increasing $v \sin i$ in steps of 5 km s$^{-1}$ and the cross-correlation with the original 
one was performed at each step. The FWHM of the CCF peak was measured and the FWHM-$v \sin i$ 
calibration was established. The $v \sin i$ values of the two components of NSVS\,1186 were derived 
from the FWHM of their CCF peaks and the aforementioned calibration relations, for few wavelength 
regions and for the best spectra. As a result the projected rotational velocity of 77$\pm$3 km s$^{-1}$ 
for the primary star and 62$\pm$4 km s$^{-1}$ for the secondary star were estimated. However, note that 
the spectral lines of the components could not be separated even at the quadratures. Thus the projected 
rotational velocities estimated in this way should involve higher uncertainties than those  given here. 

\section{Analysis}
\subsection{The Orbital Configuration}
To derive the radial velocities for the components of binary system, the 16 TFOSC spectra of the eclipsing binary  were 
cross-correlated against the spectrum of GJ\,740, a single-lined M0V star, on an order-by-order basis using the {\sc fxcor} 
package in IRAF\footnote{IRAF is distributed by the National Optical Observatory, which is operated by the Association of 
the Universities for Research in Astronomy, inc. (AURA) under cooperative agreement with the National Science 
Foundation}. The majority of the spectra showed two distinct cross-correlation peaks in the quadrature, one for each 
component of the binary. Thus, both peaks were fit independently in the quadrature with a Gaussian profile to measure 
the velocity and errors of the individual components. If the two peaks appear blended, a double Gaussian was applied to 
the combined profile using {\it de-blend} function in the task. For each of the 16 observations we then determined a 
weighted-average radial velocity for each star from all orders without significant contamination by telluric absorption 
features. Here we used as weights the inverse of the variance of the radial velocity measurements in each order, as reported 
by {\sc fxcor}. In these data, we find no evidence for a third component, since the cross-correlation function showed only 
two distinct peaks. We adopted a two-Gaussian fit algorithm to resolve cross-correlation peaks near the first and second 
quadratures when spectral lines are visible separately. Figure\,1 shows examples of cross-correlations obtained by using the 
largest FWHM at nearly first and second quadratures. The two peaks, non-blended, correspond to each component of NSVS\,1186. 
The stronger peaks in each CCF correspond to the more luminous component which has a larger weight into the observed spectrum. 

\begin{figure*}
\begin{center}
\includegraphics[width=10cm]{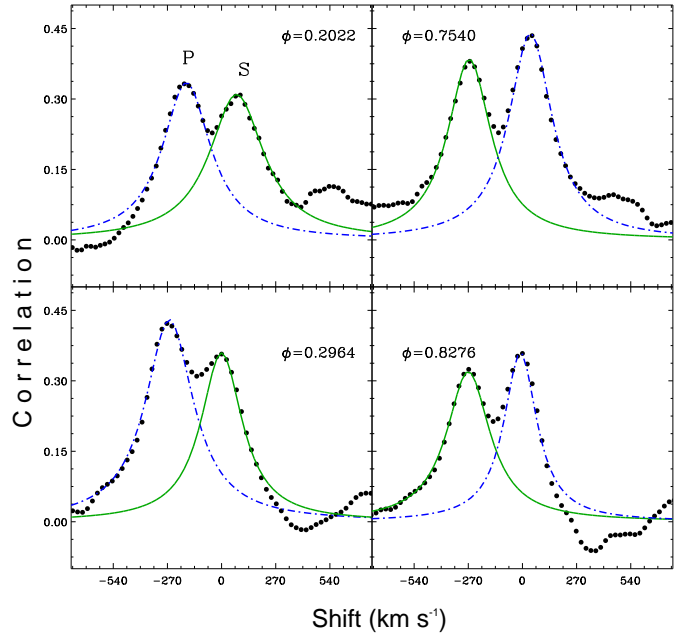}
\caption{Sample of Cross Correlation Functions (CCFs) between NSVS\,1186 and the radial velocity template 
spectrum around the first and second quadrature.}
\label{Figure 1.}
\end{center}
\end{figure*}

The heliocentric radial velocities for the primary (V$_p$) and the secondary (V$_s$) components are listed in 
Table\,2, along with the dates of observation and the corresponding orbital phases computed with the ephemeris given 
by Coughlin and Shaw (2007). The velocities in this table have been corrected to the heliocentric reference system 
by adopting a radial velocity of 9.5 km s$^{-1}$~ for the template star GJ\,740. The radial velocities listed in Table\,2 
are the weighted averages of the values obtained from the cross-correlation of orders \#4, \#5, \#6 and \#7 of 
the target spectra with the corresponding order of the standard star's spectrum. The weight $W_i = 1/\sigma_i^2$ 
has been given to each measurement. The standard errors of the weighted means have been calculated on the basis 
of the errors ($\sigma_i$) in the radial velocity values for each order according to the usual formula 
(e.g.\ Topping 1972). The $\sigma_i$ values are computed by {\sc fxcor} according to the fitted peak 
height, as described by Tonry \& Davis (1979).

First we analysed the radial velocities given in Table\,2 for the initial orbital parameters. We used the orbital period held fixed 
and computed the eccentricity of the orbit, systemic velocity and semi-amplitudes of the radial velocities. The results of 
the analysis performed with {\sc phoebe} are as follows: $e$=0.002$\pm$0.001, i.e. formally consistent with a circular orbit, $\gamma$=-80$\pm$3 km s$^{-1}$, 
$K_1$=118$\pm$5 and $K_2$=169$\pm$6 km s$^{-1}$. Using these values we estimate the projected orbital semi-major
axis and mass ratio as: $a\sin i$ =3.412$\pm$0.093 \Rsun~ and $q=\frac{M_2}{M_1}$=0.70$\pm$0.04. The observed and 
calculated radial velocities of  both components are plotted in Figure\,2 for comparison.

\begin{table*}
\begin{center}
\caption{Heliocentric radial velocities of NSVS\,1186. The columns give the heliocentric Julian date, the
orbital phase, the radial velocities of the two components with the corresponding standard deviations.
}
\label{Table 2.}
\small

\begin{tabular}{cccccc}
\hline
HJD 2400000+ & Phase & Star 1 &   &  Star 2 & \\
             &       & $V_p$ & $\sigma$  & $V_s$   	& $\sigma$	\\
\hline
55069.4412$^*$  & 0.0536 &-120.0 &14.3 &   \nodata   &  \nodata   \\		  
55069.4855  & 0.1272 &-162.0 & 9.2 &   \nodata   &  \nodata   \\		   
55069.5284  & 0.1985 &-193.6 & 4.3 &  80.6  &  4.2  \\	   
55069.5727  & 0.2721 &-198.1 & 3.9 &  85.6  &  6.3  \\	   
55068.3810  & 0.2918 &-183.7 & 3.2 &  80.0  &  8.4  \\	   
55068.4258  & 0.3663 &-171.6 & 3.0 &  47.2  &  8.6  \\	   
55068.4707  & 0.4409 &-128.7 &13.2 &   \nodata   &  \nodata   \\		  
55068.5134  & 0.5119 & -68.2 &14.9 &   \nodata   &  \nodata   \\		  
55068.5578  & 0.5857 & -28.6 &11.1 &   \nodata   &  \nodata   \\		  
55067.3674  & 0.6076 &  -6.0 & 7.3 &-184.6  & 19.7  \\		  
55068.6006  & 0.6568 &  19.2 & 3.3 &-214.3  & 12.6  \\		  
55067.4122  & 0.6820 &  28.3 & 3.1 &-235.1  & 10.2  \\		  
55067.4552  & 0.7534 &  40.0 & 1.4 &-244.3  & 13.3  \\		  
55067.4976  & 0.8240 &  19.9 & 9.2 &-230.6  &  5.2  \\	   
55067.5419  & 0.8975 &   1.1 & 2.4 &   \nodata   &  \nodata   \\		  
55067.5846$^*$  & 0.9684 & -56.5 &15.5 &   \nodata   &  \nodata   \\		  
\hline \\
\end{tabular}
\\
$^*$ These data were not used in the RV analysis. 

\end{center}
\end{table*}

\begin{figure*}
\begin{center}
\includegraphics[width=10cm]{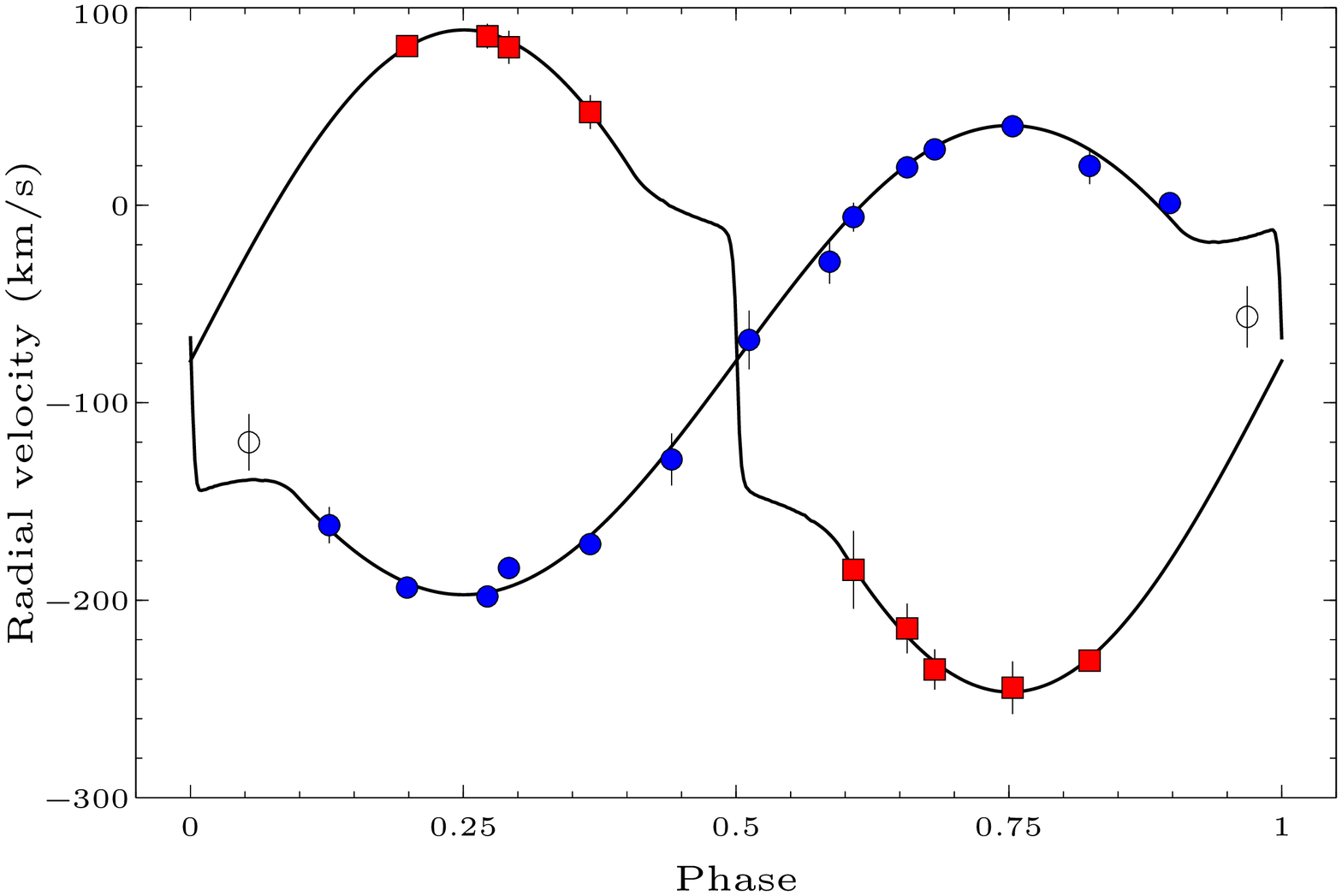}
\caption{Radial velocity curve folded on an orbital period of 0.601790 days, where phase zero is defined to 
be at primary mid-eclipse. Symbols with error bars show the RV measurements for two components 
of the system (primary:  circles, secondary: squares).  Open circles were excluded from the RV analysis.
}
\label{Figure 2.}
\end{center}
\end{figure*}

\subsection{Modelling the Light Curves}
As we mentioned in \S~1 the V-, R- and I-bandpass light curves were obtained by Coughlin and Shaw (2007). The 
differential magnitudes and orbital phases were read out carefully by the method of 
{\sc dexter}\footnote{\url{http://dc.zah.uni-heidelberg.de/sdexter}}. We used the most recent version of the 
eclipsing binary light curve modelling algorithm of Wilson \& Devinney (1971, with updates), as implemented in the {\sc phoebe} code of Pr{\v s}a \& Zwitter (2005). The code needs some input 
parameters, which depend upon the physical structures of the component stars. In the light curve solution we fixed 
some parameters whose values can be estimated from global stellar properties, such as effective temperature of the 
primary component and mass-ratio. The weighted mean of the effective temperature has been estimated to 
be  5\,250$\pm$135\,K and the mass-ratio of 0.70$\pm$0.04. Using these values we adopted the linear 
limb-darkening coefficients from Van Hamme (1993) as 0.39 and 0.28 for the primary and secondary components, 
respectively; the bolometric albedos from Lucy (1967) as 0.5, typical for a convective stellar 
envelope, the gravity brightening coefficients as 0.32 for the both components. The rotation 
of components is assumed to be synchronous with the orbital one. We start with the adjustable 
parameters: the orbital inclination, $i$, the effective temperature of the less massive star, 
T$_{eff_2}$, the potential of the components, $\Omega_1$ and $\Omega_2$, and the monochromatic 
luminosity of the more massive star, L$_{1}$. Assuming it is a detached binary we used $Mode-2$ 
of the DC program for individual solution. First we determined the preliminary elements and 
compared with the observed light curves. Since the observed light curves show distortions we
started with the two spot solution, one on the primary and other on the secondary component. The initial spot parameters 
were adopted from Coughlin \& Shaw (2007).   Best fit 
was obtained for each light curve using the same spots and the corresponding parameters are given in 
Table\,3. The parameters presented in the last column is the weighted mean of the parameters obtained 
from individual light curves. Coughlin \& Shaw (2007) analyzed only V- and R-light curves. Our analysis 
gives very different parameters when compared with those obtained by them. They estimate very low effective 
temperatures for the components with respect to their predicted masses. The fractional radii for the primary 
and secondary stars obtained by them are 15 \% and 20 \% smaller than obtained by us using the same light curves. 
In contrary, the absolute radii are 5 \% and 3 \% larger than we found.
The main difference is the locations of the spots on the surface of the stars. The spot parameters given by them for the stars should be interchanged. One spot on the south pole of the primary and one spot on the north pole of the secondary well represent the three-color light curves.  The spot coverage on the surface of the stars is about 8 \%. 

\begin{figure}
\begin{center}
\includegraphics[width=10cm]{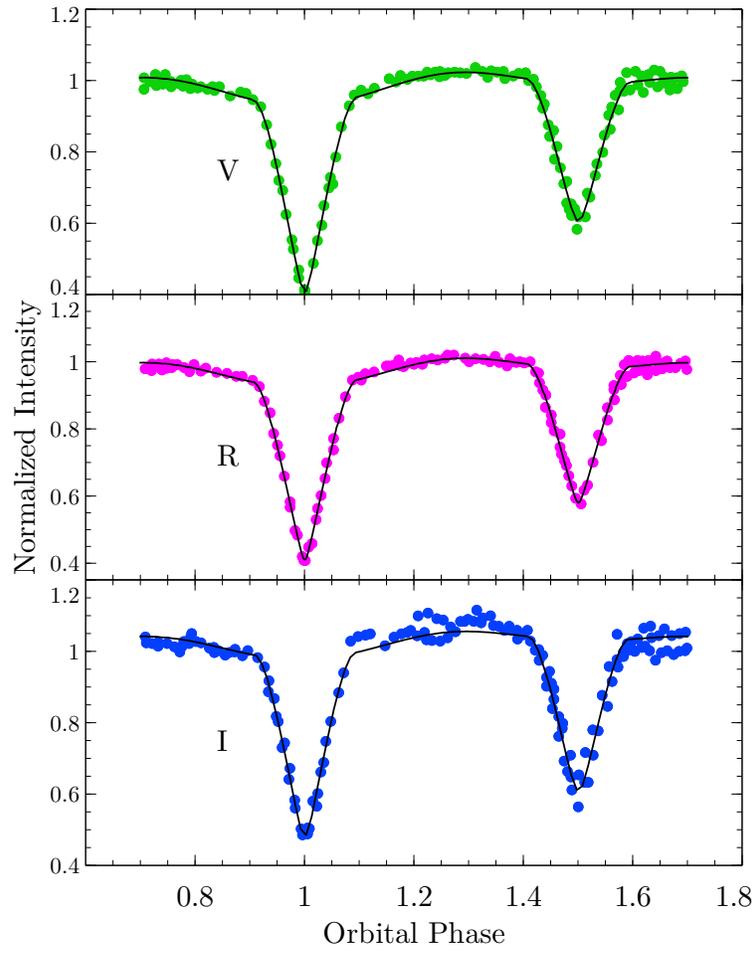}
\label{Figure 3.}
\end{center}
\caption{The phase folded V- , R- and I-bandpass light curves for NSVS\,1186. The best fitting solutions 
represented by the solid lines are also plotted for comparison (see text).}
\end{figure}

\begin{table*}
\begin{center}
\caption{Results of the V-, R- and I-bandpass light curve analysis for NSVS\,1186. The adopted values are 
the weighted means of the values determined from the individual light curves.}
\begin{tabular}{lcccc}
\hline
Parameters &V &R  &I&Adopted  \\
\hline
$i^{o}$			               			&87.0$\pm$0.5	 & 88.6$\pm$0.2		& 88.3$\pm$0.2	 	& 88.0$\pm$0.4	\\
T$_{eff_1}$ (K)							&5\,250[Fix]	 & 5\,250[Fix]		& 5\,250[Fix]	 	& 5\,250[Fix]	\\
T$_{eff_2}$ (K)							&5\,040$\pm$12	 & 4\,990$\pm$14	& 5\,110$\pm$34	 	& 5\,020$\pm$16	\\
$\Omega_1$								&4.242$\pm$0.049 & 4.274$\pm$0.049  & 4.260$\pm$0.072	& 4.262$\pm$0.052\\
$\Omega_2$								&3.810$\pm$0.026 & 3.739$\pm$0.020	& 3.989$\pm$0.054   & 3.790$\pm$0.026\\				
r$_1$									&0.289$\pm$0.004 & 0.288$\pm$0.004	& 0.286$\pm$0.006	& 0.288$\pm$0.004\\
r$_2$									&0.263$\pm$0.003 & 0.267$\pm$0.002	& 0.250$\pm$0.005	& 0.264$\pm$0.003\\
${L_{1}}/{(L_{1}+L_{2})}$ 				&0.604$\pm$0.009 & 0.586$\pm$0.009	& 0.591$\pm$0.013	&   \nodata 	\\
$\sum(W(O-C)^2) $	     							&0.041			 & 0.026			& 0.131				&   \nodata	\\
weight							&3			 & 5			& 1				&   \nodata	\\		
\hline
\end{tabular}
\end{center}
\end{table*}

\subsection{Absolute Parameters}
Combining the parameters obtained by radial velocity and light curve analyses we calculated the physical 
parameters of the components as listed in Table\,4. We compare the locations of the components in the 
effective temperature-luminosity diagram, i.e. Hertzsprung-Russell diagram (Figure\,4). While the location of the 
more massive primary star agrees well with a star of having the mass, the secondary component 
appears to be more luminous, more probably originating larger effective temperature and radius. 
We also computed the  radii of the components as  R$_1$=0.983$\pm$0.030\,\Rsun~ 
and R$_2$=0.901$\pm$0.026\,\Rsun~ for the primary and secondary stars, respectively.

We computed the luminosities of the components using these radii and effective temperatures  as 
L$_1$ =0.649$\pm$0.069\,\Lsun~ and L$_2$=0.456$\pm$0.049\,\Lsun~ for the primary and 
secondary stars, respectively. We compare the positions of the components in L-T$_e$ diagram 
with evolutionary tracks of  pre- and post- main sequence evolution and as well as with the ZAMS, as shown in Figure\,4.
For constructing solar metallicity ZAMS and evolutionary tracks, we used the Cambridge version of the  the {\sc stars}
code which was originally developed by Eggleton (1971) and had substantial revision up to date 
(Eldridge \& Tout, 2004). The current version is also capable to account molecular contribution on equation of state 
according to Marigo (2002) approximation which may be crucial for low mass stars.
The less massive component seems to hotter and more luminous with respect to the models.
The physical parameters determined by us 
are very different from those obtained by Coughlin and Shaw (2007). While they estimate effective 
temperatures between 4\,370 and 5\,250 K from the observed colours they adopted an effective 
temperature of 3\,970 K for the more massive star. The masses of the components estimated to be 
0.94 and 0.87 and radii of 1.03 and 0.93, and an effective temperature of 3\,750 K for the secondary 
star. The locations of the component stars in the HR diagram correspond to stars having masses below  
0.4\,\Msun. The J-,H-, and K- band extinctions in the direction of the variable 
taken from  Schlegel et al. (1998)  as 
A$_J$=0.046, A$_H$=0.030 and A$_K$=0.019 magnitudes. We computed the absolute magnitudes as 
M$_J$=3.96$\pm$0.10,  M$_H$=3.50$\pm$0.07 and  M$_K$=3.46$\pm$0.07 for the primary star.  Then using the JHK 
magnitudes given in \S~2.1.1 we estimate the distance to the variable as 668$\pm$25, 689$\pm$24 
and 678$\pm$22 parsec in J,H, and K bands, respectively.

\begin{table*}
\begin{center}
\setlength{\tabcolsep}{2pt} 
\caption{Fundamental parameters of the system.}
  \label{parameters}
  \begin{tabular}{lcc}
  \hline
   Parameter 						& Primary	&	Secondary	\\
   \hline
   Mass (M$_{\odot}$) 				& 0.870$\pm$0.074  & 0.607$\pm$0.053	\\
   Radius (R$_{\odot}$) 				& 0.983$\pm$0.030  & 0.901$\pm$0.026	\\
   $\log~g$ ($cgs$) 				& 4.392$\pm$0.019  & 4.311$\pm$0.021	\\
   $T_{eff}$ (K)					& 5\,250$\pm$135	&5\,020$\pm$135 		\\
   $\log~(\rm{L/L_{\odot}})$				&-0.188$\pm$0.045	& -0.341$\pm$0.046  		\\
   $(v \sin~i)_{obs.}$ (km s$^{-1}$)			& 77$\pm$3 	& 62$\pm$4	 	\\
   $(v \sin~i)_{calc.}$ (km s$^{-1}$)			& 83$\pm$3 	& 76$\pm$2 	\\
\hline  
  \end{tabular}
\medskip\\

\end{center}\end{table*}

\begin{figure*}
\begin{center}
\includegraphics[width=10cm]{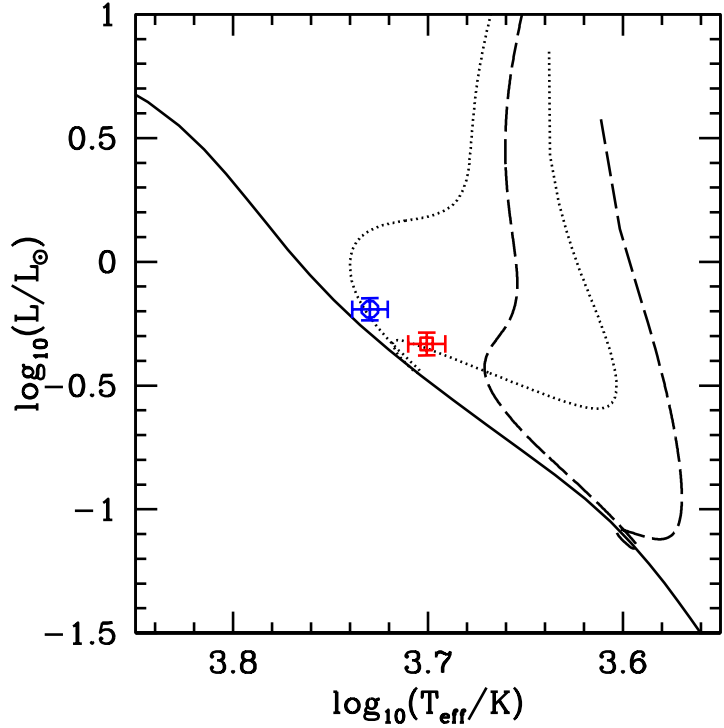}
\caption{Comparison between evolutionary models and the physical parameters of NSVS 1186 in the effective 
temperature-luminosity diagram where the luminosities determined from the light and radial velocity 
curves. Theoretical evolutionary tracks, both pre- and post- main sequence,  for masses of  0.870 \Msun~ (dotted line) and 0.607 \Msun~ (dashed line) and ZAMS (solid line) with solar abundance 
are calculated with the {\sc stars} code. The circle and square denote the primary and secondary 
stars. }
\label{Figure X.}
\end{center}
\end{figure*}

\section{Oversized Stars in the Low-Mass Eclipsing Binaries}
Abt (1963) and Duquennoy, Mayor and Halbwachs (1991) report that binary stars are more common 
than single stars at masses above that of the Sun. In contrary, Reid \& Gizis (1997) and Delfosse 
et al. (1999) suggest that binaries are not very common in low-mass stars, although approximately 
75 \% of all stars in our Galaxy are low-mass dwarfs with masses smaller than 0.7 \Msun. Due to the 
low binary fractions and their faintness very few low-mass eclipsing binary systems have been observed 
so far both photometrically and spectroscopically, yielding accurate physical parameters. 
Recently in a pioneer study, Ribas (2003,2006) collected the available masses and radii of the low-mass stars
and compared with those obtained by stellar evolution models. Even small numbers of the sample, a total of eight 
double-lined eclipsing binaries whose mass and radius are determined with an accuracy of better than 3 \%, 
the comparison evidently revealed that the observed radii are systematically larger than the models. However, 
the effective temperatures are cooler than the theoretical calculations, being the luminosities in agreement 
with those of single stars with the same mass. This discrepancy between the models and observations has been explained 
by Mullan and MacDonald (2001), Torres et al. (2006), Ribas (2006) and Lopez-Morales (2007) and others by the high 
level magnetic activity in the low-mass stars. 
They suggest that stellar activity may be responsible for the observed discrepancy through inhibition of 
convection or effects of a significant spot coverage. However, Berger et al. (2006) compared the interferometric 
radii of low-mass stars with those estimated from theory and find a correlation between an increase in 
metallicity and larger-than-expected radius, i.e. the larger radii are caused by differences in metallicity. 

The number of double-lined low-mass eclipsing binaries having precise photometric and spectroscopic observations 
is reached to 21  according to knowledge of the authors. Combining the results of the analysis of both photometric and spectroscopic observations 
accurate masses, radii, effective temperatures and luminosities  of the components have been obtained. 
In Table\,5 we list absolute parameters for the low-mass stars with their standard deviations. 
In Figure\,5 we plot the radii of the low-mass stars in the eclipsing binary systems as a function of mass. 
Theoretical mass-radius (M-R) diagrams for the zero-age main-sequence stars with solar abundance taken 
from the {\sc stars} code and Siess et al. (2000) are also plotted for comparison.

The stars below 0.3 \Msun~ appear to trace the theoretical M-R diagram. However, larger mass stars, 
above 0.3 \Msun~, significantly deviate from the theoretical M-R digram.  We also computed projected
 rotational velocities of the components using their radii 
and orbital period. Since the orbital periods are generally short we assume spin-orbit synchronization. 
The rotational velocities are plotted in Figure\,6 as a function of the radius difference calculated as
$(R_{obs}-R_{zams})/R_{zams}$. This plot shows that as the rotational 
velocity increases, the difference in  radius, gets larger for the stars given in Table\,5.

A low-mass star has a convective atmosphere with a radiative core. The depth of the convective zone is about 
0.28 R for a star of mass about 0.9 \Msun,~ and gradually increases to about 0.41 R for a star with a 
mass about 0.4 \Msun.  Stars are thought to be fully convective below about 0.35 \Msun.~ Therefore, the internal 
structure of such a low-mass star with a deep convective zone is tightly dependent on the mixing length 
parameter, $\alpha$. In Figure\,7a and b we compare locations of the low-mass stars in the M-R and 
M-T$_{eff}$ diagrams with the theoretical calculations for various $\alpha$ parameters. Convection is modelled
by mixing length theory (B{\"o}hm-Vitense 1958) with the ratio of mixing length to pressure scale height
(i.e. $\alpha = l/H_p$). Theoretical 
models show that there is no significant separation in the  M-R relation for the stars below about 
0.6 \Msun~, depending on the mixing length parameter. For masses above this value a separation is 
revealed. However, as the $\alpha$ parameter 
increases the computed effective temperature gets higher. A separation is clearly seen even for the 
masses below 0.6 \Msun~. We also plotted empirical parameters of the low-mass stars in these panels. This 
comparison clearly exposes that the discrepancies in radii and effective temperatures could not be explained only by  
changing $\alpha$ parameter as it is obvious in M-R and M-T$_{eff}$ plots represented in Figure\,7.
 On the other hand Demory et al. (2009) showed that there is no significant correlation between metallicity
and radius of the single, low-mass stars. An alternative explanation remains to be the magnetic 
activity responsible for the observed larger radii 
but cooler effective temperatures. As demonstrated by D'Antona et al. (2000), Mullan and MacDonald (2001) and 
Chabrier et al.(2007) magnetic fields change the 
evolution of low-mass stars. Due to the high magnetic activity in the fast-rotating dwarfs their surface 
are covered by dark spots. Spot coverage in active dwarfs yields larger radii and lower effective temperatures.

\begin{table*}
\scriptsize
\begin{center}
 \setlength{\tabcolsep}{2.5pt} 
  \caption{The mass, radius, effective temperature and orbital period of the 
  low-mass stars in the double-lined eclipsing binaries. }
  \label{parameters}
\begin{tabular}{lcrclrclrclccr}
\hline
Object	&Star	&\multicolumn{3}{c}{Mass/\Msun}	&\multicolumn{3}{c}{Radius/\Rsun}&\multicolumn{3}{c}{T$_{\rm eff}$/K}&P$_{\rm orb}/\rm day$	&Refs	&\\
\hline
CM Dra	&A	&0.2310	&$\pm$	&0.0009	&0.2534	&$\pm$	&0.0019	&3130	&$\pm$	&70	&1.276	&1	&\\
	&B	&0.2141	&$\pm$	&0.0008	&0.2398	&$\pm$	&0.0018	&3120	&$\pm$	&70	&	&	&\\
YY Gem	&A	&0.5992	&$\pm$	&0.0047	&0.6194	&$\pm$	&0.0057	&3820	&$\pm$	&100	&0.820	&2	&\\
	&B	&0.5992	&$\pm$	&0.0047	&0.6194	&$\pm$	&0.0057	&3820	&$\pm$	&100	&	&	&\\
CU Cnc	&A	&0.4349	&$\pm$	&0.0012	&0.4323	&$\pm$	&0.0055	&3160	&$\pm$	&150	&2.794	&3	&\\
	&B	&0.3992	&$\pm$	&0.0009	&0.3916	&$\pm$	&0.0094	&3125	&$\pm$	&150	&	&	&\\
GU Boo	&A	&0.6101	&$\pm$	&0.0064	&0.6270	&$\pm$	&0.0160	&3920	&$\pm$	&130	&0.492	&4	&\\
	&B	&0.5995	&$\pm$	&0.0064	&0.6240	&$\pm$	&0.0160	&3810	&$\pm$	&130	&	&	&\\
TrES-Her0-07621	&A	&0.4930	&$\pm$	&0.0030	&0.4530	&$\pm$	&0.0600	&3500	&$\pm$	& \nodata &1.137	&5	&\\
	&B	&0.4890	&$\pm$	&0.0030	&0.4520	&$\pm$	&0.0500	&3395	&$\pm$	& \nodata	&	&	&\\
2MASS J05162881+2607387	&A	&0.7870	&$\pm$	&0.0120	&0.7880	&$\pm$	&0.0150	&4200	&$\pm$	& \nodata	&2.619	&6	&\\
	&B	&0.7700	&$\pm$	&0.0090	&0.8170	&$\pm$	&0.0100	&4154	&$\pm$	& \nodata	&	&	&\\
UNSW-TR 2	&A	&0.5290	&$\pm$	&0.0350	&0.6410	&$\pm$	&0.0500	& \nodata	&$\pm$	& \nodata	&2.144	&7	&\\
	&B	&0.5120	&$\pm$	&0.0350	&0.6080	&$\pm$	&0.0600	& \nodata	&$\pm$	& \nodata	&	&	&\\
NSVS 06507557	&A	&0.6560	&$\pm$	&0.0860	&0.6000	&$\pm$	&0.0300	&3960	&$\pm$	&80	&0.520	&8	&\\
	&B	&0.2790	&$\pm$	&0.0450	&0.4420	&$\pm$	&0.0240	&3365	&$\pm$	&80	&	&	&\\
NSVS 02502726	&A	&0.7140	&$\pm$	&0.0190	&0.6740	&$\pm$	&0.0600	&4300	&$\pm$	&200	&0.560	&9	&\\
	&B	&0.3470	&$\pm$	&0.0120	&0.7630	&$\pm$	&0.0050	&3620	&$\pm$	&205	&	&	&\\
T-Lyr1-17236	&A	&0.6795	&$\pm$	&0.0107	&0.6340	&$\pm$	&0.0430	&4150	&$\pm$	& \nodata	&8.430	&10	&\\
	&B	&0.5226	&$\pm$	&0.0061	&0.5250	&$\pm$	&0.0520	&3700	&$\pm$	& \nodata	&	&	&\\
2MASS J01542930+0053266	&A	&0.6590	&$\pm$	&0.0310	&0.6390	&$\pm$	&0.0830	&3730	&$\pm$	&100	&2.639	&11	&\\
	&B	&0.6190	&$\pm$	&0.0280	&0.6100	&$\pm$	&0.0930	&3532	&$\pm$	&100	&	&	&\\
GJ 3236	&A	&0.3760	&$\pm$	&0.0170	&0.3828	&$\pm$	&0.0072	&3310	&$\pm$	&110	&0.770	&12	&\\
	&B	&0.2810	&$\pm$	&0.0150	&0.2992	&$\pm$	&0.0075	&3240	&$\pm$	&110	&	&	&\\
SDSS-MEB-1	&A	&0.2720	&$\pm$	&0.0200	&0.2680	&$\pm$	&0.0090	&3320	&$\pm$	&130	&0.410	&13	&\\
	&B	&0.2400	&$\pm$	&0.0220	&0.2480	&$\pm$	&0.0080	&3300	&$\pm$	&130	&	&	&\\
BD -22 5866	&A	&0.5881	&$\pm$	&0.0029	&0.6140	&$\pm$	&0.0450	& \nodata	&$\pm$	& \nodata	&2.211	&14	&\\
	&B	&0.5881	&$\pm$	&0.0029	&0.5980	&$\pm$	&0.0450	& \nodata	&$\pm$	& \nodata	&	&	&\\
NSVS 01031772	&A	&0.5428	&$\pm$	&0.0027	&0.5260	&$\pm$	&0.0028	&3615	&$\pm$	&72	&0.368	&15	&\\
	&B	&0.4982	&$\pm$	&0.0025	&0.5088	&$\pm$	&0.0030	&3513	&$\pm$	&31	&	&	&\\
NSVS 11868841	&A	&0.8700	&$\pm$	&0.0740	&0.9830	&$\pm$	&0.0300	&5260	&$\pm$	&110	&0.602	&This paper	&\\
	&B	&0.6070	&$\pm$	&0.0530	&0.9010	&$\pm$	&0.0260	&5020	&$\pm$	&110	&	&	&\\
GJ 2069A	&A	&0.4329	&$\pm$	&0.0018	&0.4900	&$\pm$	&0.0800	& \nodata	&$\pm$	& \nodata	&2.771	&16	&\\
	&B	&0.3975	&$\pm$	&0.0015	&0.3300	&$\pm$	&0.0400	& \nodata	&$\pm$	& \nodata	&	&	&\\
2MASS J04463285+1901432	&A	&0.4700	&$\pm$	&0.0500	&0.5700	&$\pm$	&0.0200	&3320	&$\pm$	&150	&0.630	&18	&\\
	&B	&0.1900	&$\pm$	&0.0200	&0.2100	&$\pm$	&0.0100	&2910	&$\pm$	&150	&	&	&\\
NSVS 6550671	&A	&0.5100	&$\pm$	&0.0200	&0.5500	&$\pm$	&0.0100	&3730	&$\pm$	&60	&0.193	&19	&\\
	&B	&0.2600	&$\pm$	&0.0200	&0.2900	&$\pm$	&0.0100	&3120	&$\pm$	&65	&	&	&\\
IM Vir	&A	&0.9810	&$\pm$	&0.0120	&1.0610	&$\pm$	&0.0160	&5570	&$\pm$	&100	&1.309	&20	&\\
	&B	&0.6644	&$\pm$	&0.0048	&0.6810	&$\pm$	&0.0130	&4250	&$\pm$	&130	&	&	&\\
RXJ0239.1	&A	&0.7300	&$\pm$	&0.0090	&0.7410	&$\pm$	&0.0040	&4645	&$\pm$	&20	&2.072	&21	&\\
	&B	&0.6930	&$\pm$	&0.0060	&0.7030	&$\pm$	&0.0020	&4275&$\pm$	&	15&	&	&\\

\hline
\end{tabular}
\begin{list}{}{  }
\item[Ref:]{ (1)Morales et al. (2009a), (2)Torres \& Ribas (2002), 
(3)Ribas (2003), (4)L{\'o}pez-Morales \& Ribas (2005),  (5)Creevey et al. 
(2005),  (6)Bayless \& Orosz (2006), (7)Young et al. (2006),  (8){\c C}ak{\i}rl{\i} \& 
Ibano{\v g}lu (2010), (9){\c C}ak{\i}rl{\i}, Ibano{\v g}lu \& G{\"u}ng{\"o}r (2009),
(10) Devor et al. (2008), (11)Becker et al. (2008),  (12)Irwin et al. (2009),  (13)Blake et al. (2008),
(14)Shkolnik et al. (2008), (15)L{\'o}pez-Morales et al. (2006), (16)Delfosse et al. (1999),
(17)Beatty et al. (2007), (18)Hebb et al. (2006), (19)Dimitrov  \& Kjurkchieva (2010)  ,(20)Morales et al. (2009b), (21)L{\'o}pez-Morales \& Shaw 
(2007)}
\end{list}
\end{center}\end{table*}

\begin{figure}
\begin{center}
\includegraphics[width=10cm]{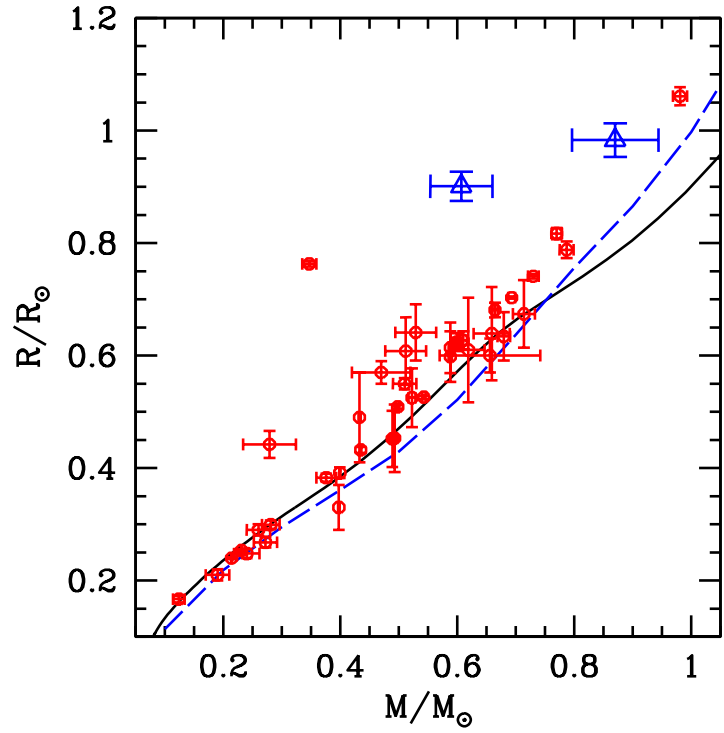}
\caption{Locations of the low-mass stars in the mass-radius diagram. The components of NSVS\,1186 are shown by triangles. The solid and dashed lines represent the ZAMS models 
taken from the {\sc stars} code  and Siess et al.(2000), respectively, with solar abundance.}
\label{Figure x.}
\end{center}\end{figure}

\section{Discussion}
We obtained the radial velocity curves for the low-mass eclipsing binary NSVS\,1186. We re-analysed  the existing V-, R- 
and I-bandpass light curves. Since the light curves are distorted they are modelled assuming one spot located on each star. 
Then combining the photometric and spectroscopic solutions we derived the absolute physical parameters for 
both components. The standard deviations of the parameters have been determined using the {\sc jktabsdim} 
code (Southworth and Clausen 2005). The less massive component appears to have a larger radius and higher temperature
 with respect to the its mass. On the other hand the primary star seems to have a larger radius but the effective temperature is in
 agreement with that expected from its mass.  We suppose that the larger radii of both components  
may be originated from a higher activity in the system at the time that photometry is performed. The locations 
of the components in the H-R diagram have been plotted and compared with theoretical models. The secondary 
component seems to have slightly brighter with respect to its mass.

Basic physical parameters determined from double lined  mass binaries , such as mass, radius, effective temperature and orbital 
period, for low-mass stars are collected. The locations of the low-mass stars in the mass-radius panel are compared 
with the theoretical ZAMS models. Since the low-mass stars evolve very slowly, we prefer a comparison with the ZAMS 
models. Theoretical models tend to underestimate the radii at a given mass in the range of 0.3 and 0.6 \Msun~. Due to 
the evolutionary effect on the radius, we take into account only this mass range, preventing any confusion. Since 
low-mass stars have deeper convective layers in their  atmospheres we also computed the radii and temperatures for 
the above-mentioned mass range for various mixing-length parameters. We find that mixing-length parameter does 
not significantly affect the radii, in contrary, it changes considerably the effective temperatures of the 
stars. Comparison with the observations shows that the observed larger radii of the late K and M dwarfs can 
not only be  represented with the values of mixing-length parameters. These comparisons confirm that magnetic field 
induced inhibition of convection in fast-rotating low-mass stars with deep convective zones lead to larger radii 
and lower effective temperatures than those expected from theoretical models (D'Antona et al. 2000, Mullan 
and MacDonald 2001, Chabrier et al. 2007) .  

\begin{figure}
\begin{center}
\includegraphics[width=10cm]{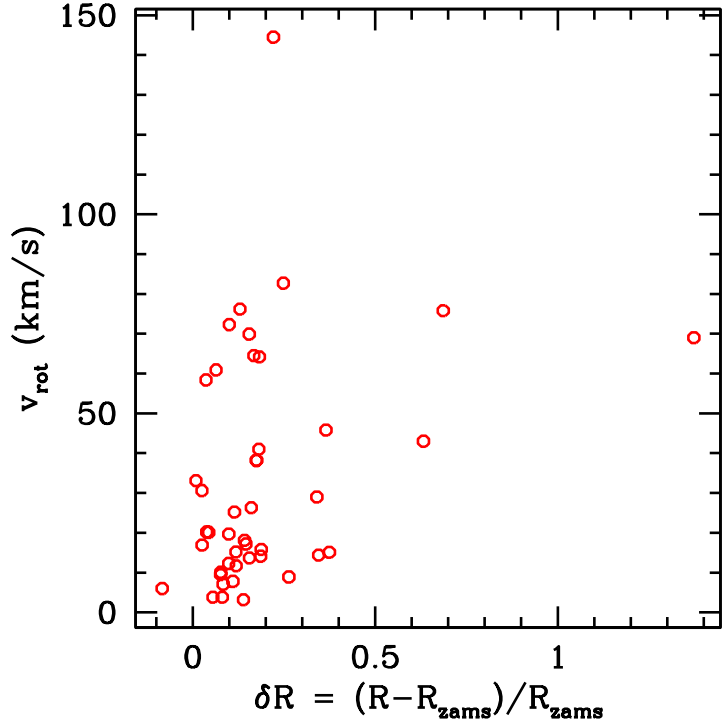}
\caption{The rotational velocities are plotted against the radius difference from the  ZAMS with the same mass.}
\label{Figure y.}
\end{center}
\end{figure}

\begin{figure}[]
\begin{center}
\includegraphics[width=10cm]{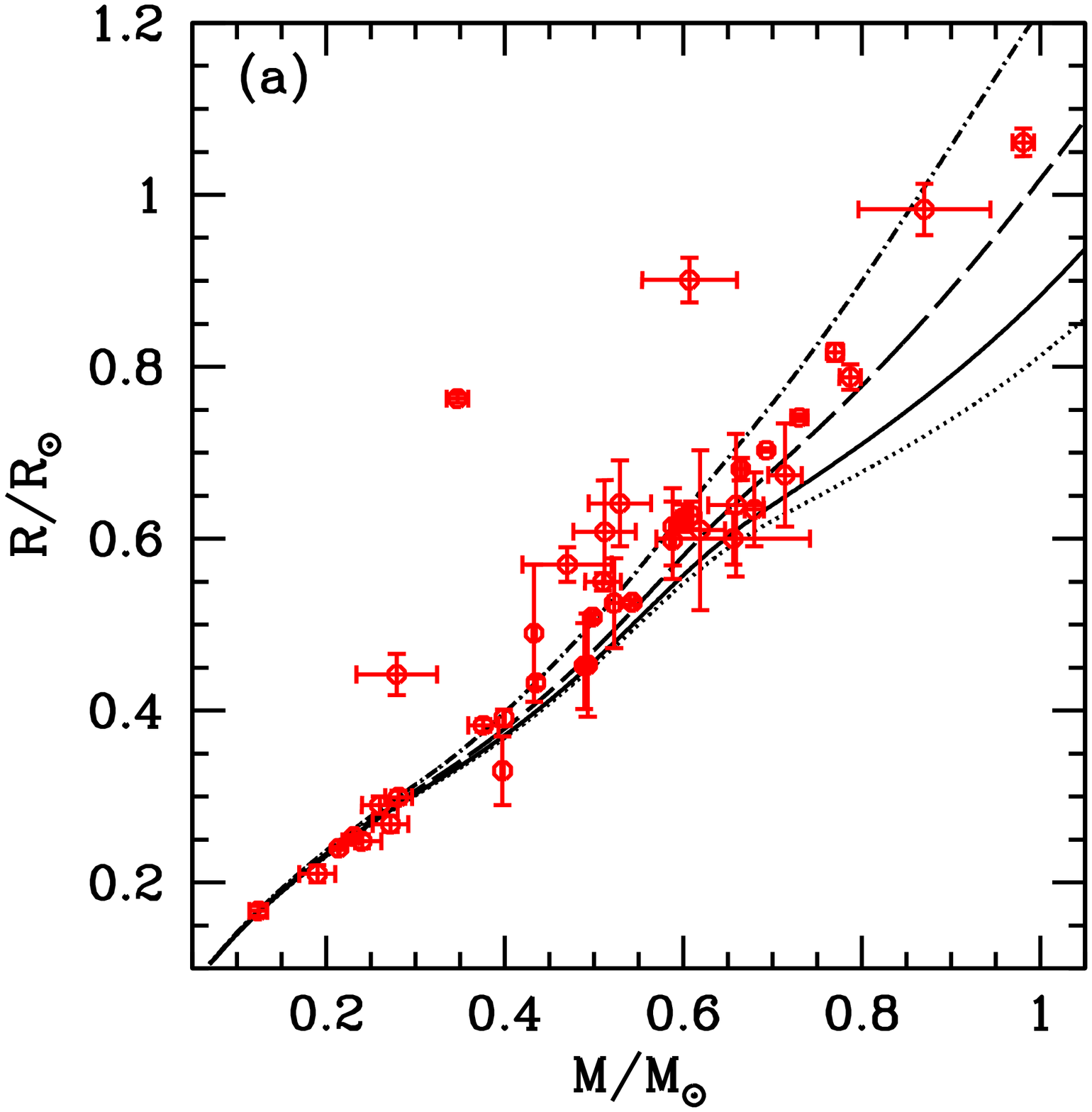}
\includegraphics[width=10cm]{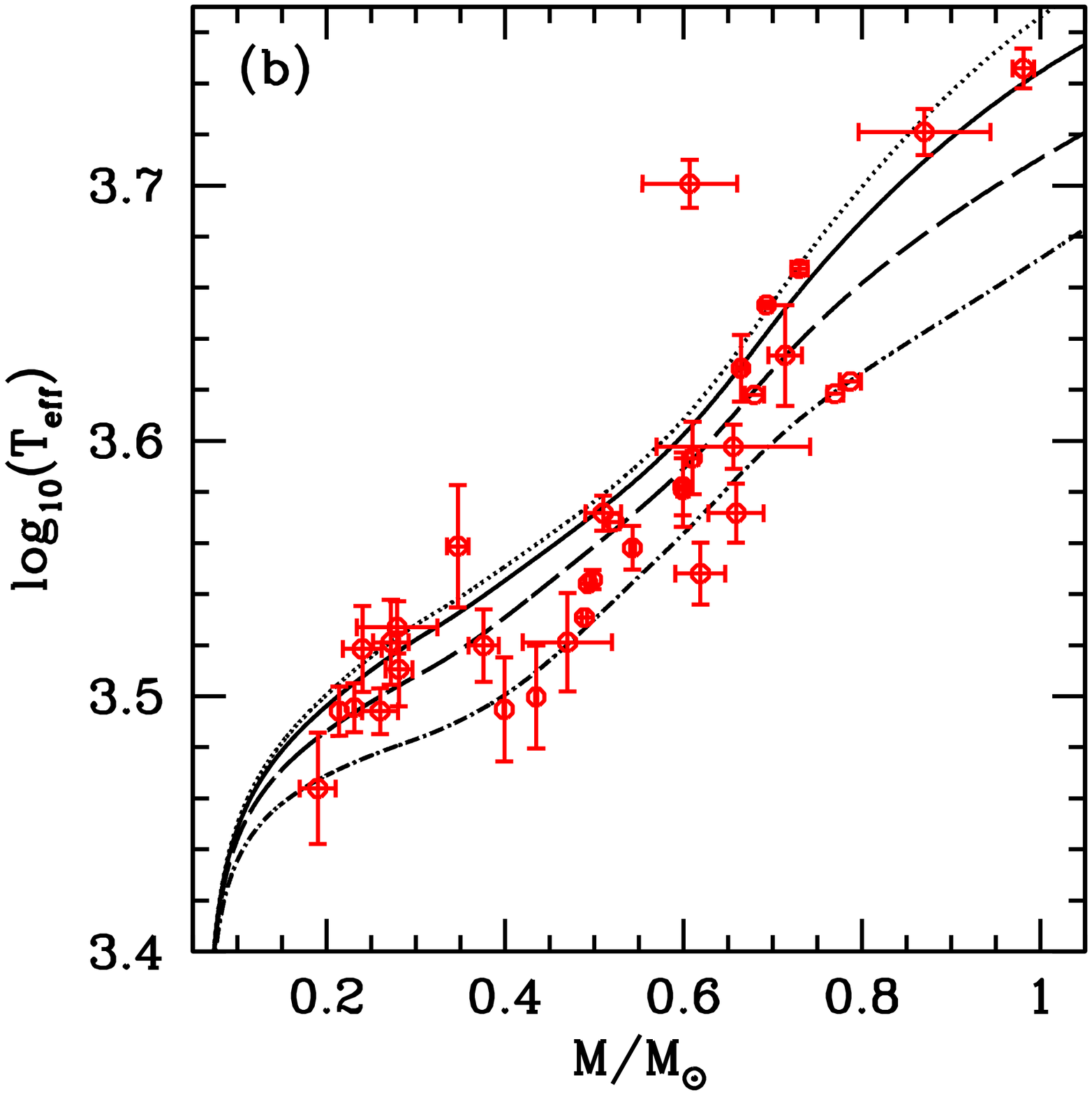}
\caption{Mass-radius (a) and mass-effective temperature (b) relationships for zero-age main-sequence low-mass  stars  calculated with the {\sc stars} code for  
values of the mixing length parameter as $\alpha=4$ (dotted line), $\alpha=2$ (solid line), $\alpha=1$ (dashed line) and 
$\alpha=0.5$ (dot-dashed line). The values of the alpha parameter higher than 0.5 appear to effect the radius slightly for a star 
below 0.7 \Msun. As the alpha increases the effective temperatures get higher.}
\label{Figure z.}
\end{center}
\end{figure}

\section*{Acknowledgments} 
The authors acknowledge generous allotments of observing time at TUBITAK National Observatory (TUG) of Turkey. We also 
wish to thank the Turkish Scientific and Technical Research Council for supporting this work through grant Nr. 108T210 
and  EB{\.I}LTEM Ege University Science Foundation Project No:08/B\.{I}L/027 . Dr. Esin Sipahi is acknowledged for the aid during the computation. We would like express our sincere gratitudes to the anonymous referee whose suggestions made the paper very clearer.

\end{document}